\documentclass{statsoc}

\usepackage[utf8x]{inputenc}

\usepackage[english]{babel}
\usepackage[utf8x]{inputenc}
\usepackage[T1]{fontenc}

\usepackage[a4paper]{geometry}

\usepackage{amsmath}
\usepackage{graphicx}
\usepackage[colorinlistoftodos]{todonotes}
\usepackage[colorlinks=true, allcolors=blue]{hyperref}

\newcommand{\edit}[1]{\textcolor{black}{#1}}
\newcommand{\secondedit}[1]{\textcolor{black}{#1}}

\usepackage{natbib}

\title[A Comparison of Prior Elicitation Aggregations]{A Comparison of Prior Elicitation Aggregation using the Classical Method and SHELF}

\author[C.J. Williams, K.J. Wilson and N. Wilson]{Cameron J. Williams}
\address{Newcastle University,
Newcastle Upon Tyne,
United Kingdom.}
\email{cameron.williams@ncl.ac.uk}
\author[C.J. Williams, K.J. Wilson and N. Wilson]{Kevin J. Wilson}
\address{Newcastle University,
Newcastle Upon Tyne,
United Kingdom}
\author[C.J. Williams, K.J. Wilson and N. Wilson]{Nina Wilson}
\address{Newcastle University,
Newcastle Upon Tyne,
United Kingdom}

\begin{document}
\maketitle

\begin{abstract}
Subjective Bayesian prior distributions elicited from experts can be aggregated together to form group priors. This paper compares aggregated priors formed by Equal Weight Aggregation, the Classical Method, and the Sheffield Elicitation Framework to each other and individual expert priors, using an expert elicitation carried out for a clinical trial. Aggregation methods and individual expert prior distributions are compared using proper scoring rules to compare the informativeness and calibration of the distributions. The three aggregation methods outperform the individual experts, and the Sheffield Elicitation Framework performs best amongst them.
\end{abstract}



\section{Introduction}

Bayesian priors have long been used to incorporate expert knowledge into modelling and analysis. To ensure the prior reflects the wider state of knowledge in the field, multiple experts can be consulted. This leads to the issue of how to represent the views of multiple experts in a single analysis. Each expert's beliefs could be represented by their own individual prior, and the analysis completed as many times as there are experts. Alternatively, the experts' beliefs could be aggregated to form a group prior, allowing for a single analysis \citep{West1984}. Group priors have been shown to be more informative and have better probabilistic calibration than priors of individual experts \citep{Clemen2008,Lin2009}.

There are many methods of aggregating a number of expert prior distributions into a single group prior. Broadly speaking, the two widely used methods for aggregation are either mathematical rule based, where individual priors are combined according to a predetermined mathematical rule, or behavioural, where experts work together to form a group prior. This paper investigates the most widely used rule based method, the Classical Method (CM), and behavioural method, the Sheffield Elicitation Framework (SHELF), to compare the performance of the two during the same elicitation. These methods are discussed in detail in \citet{EFSA2014}.

While each have methodological differences that provide benefits in different circumstances, their statistical performance relative to each other has not previously been directly compared. In addition to these two common aggregation methods, we also compare them against an Equal Weight (EW) aggregation and the individual experts as a baseline. An elicitation for a clinical trial into diagnostic tests for Motor Neuron Disease (MND) was used to form the range of priors.

The performance of the CM has previously been compared to EWs and the single best expert.  \cite{Clemen2008} found that aggregating with the CM led to more informative, but less well calibrated, distributions compared to EWs. Compared to both methods, using a single best expert gave a less informative and calibrated distribution. \cite{Lin2009} found that the CM and EWs lead to similar performance by the decision maker, and that both outperformed a single best expert.

\cite{Cooke2008} investigated a weighting scheme using the number of papers an expert has published which had been cited by other experts involved in the elicitation. The CM was shown to outperform this weighting method.

This paper will compare the CM and SHELF, using an elicitation conducted to inform the sample size calculation of a clinical trial. Section 2 provides a background and summary of the methods used to aggregate individual expert priors into a single group prior. Section 3 outlines the trial, elicitation and resulting distributions. Section 4 provides a discussion and comparison between the methods for the elicited distributions for the trial, and for additional validation questions for which the answers were known to the elicitors. We provide a short conclusion in Section 5.

\section{Methods}

Prior distributions can be formed using information from a variety of sources. In many cases, these sources can be yet unquantified expert knowledge. Probability distributions can be elicited in order to transform this knowledge into a form useful for statistical analysis. This can pose a challenge, as experts do not tend to think in terms of probability distributions. Thus, their views need to be converted to this form.

An elicitation session aims to construct probability distributions that accurately represent an expert's views. The session is run by an elicitor. The design of the elicitation involves several steps \citep{Choy2009}. The reasons and motivation for utilising prior knowledge are determined, to ensure whatever information is elicited is fit for purpose. This also forms the starting point for investigating the availability of expert knowledge, and the development of a statistical model. 

The prior distributions can then be elicited using a number of different methods, and from a single or multiple experts. This section will review some of the common ways to elicit from groups of experts, focusing on those we will compare in later sections. 



\subsection{Mathematical Aggregation Methods}

Mathematical aggregation methods rely on assigning weights to each expert's prior distributions, according to a predefined rule. For elicited priors from $n$ experts, each expert $i$ = 1, \dots, $n$, has an individual prior $f_i(\theta)$ for the parameter of interest, $\theta$. A weight for each expert, $w_i$, can then to be used to calculate an aggregated group prior, $f(\theta).$ Note that $\sum^n_{i=1} w_i = 1.$

In a simple linear pooling case, the individual priors are multiplied by the experts' weights, and summed \citep{OHagan2006}. This method forms a prior where the value of $f(\theta)$ is the weighted arithmetic mean of each expert's prior distribution.

\begin{equation}
f(\theta) = \sum^n_{i=1} w_i f_i(\theta)
\end{equation}

Alternatively, a log linear pooling method can be used. The group prior in this case is a weighted geometric mean of the individual experts' priors, \edit{with a normalising constant $k$ to ensure the density $f(\theta)$ integrates to one}.

\begin{equation}
f(\theta) = k \prod^n_{i=1} f_i(\theta)^{w_i}
\end{equation}





For the purposes of the comparisons in this paper, linear pooling will be used. There are a number of potential issues that can arise under a log linear pooling scheme.

Firstly, if any expert provides a probability of zero for a particular value of $\theta$, i.e. $f_i(\theta^*) = 0$, the aggregated distribution will have a probability of zero for  $\theta^*$. This can be the case when a minimum and maximum value of $\theta$ is elicited, and ensures the largest minimum and smallest maximum become the bounds of the group distribution. This can be problematic, especially \secondedit{when one expert gives bounds} vastly different to the other experts, as the final prior distribution may completely ignore large proportions of the densities provided by the majority of experts. Furthermore, in the case where two experts' distributions do not overlap at all, the calculated group prior would have no density.


Additionally, in cases where there is very little overlap between the densities of the experts' priors, log linear pooling can result in an aggregated distribution that places the majority of its density in the region between expert priors. This results in a distribution which places a high probability on values of $\theta$ that no individual expert has deemed likely, and may not have a realistic practical interpretation. 


\subsubsection{Equal Weights Aggregation}

The most straightforward method of assigning weights to experts is to assign each an Equal Weight (EW). The weights of all experts are assigned so they will sum to one, with

\begin{equation}
    w_i = \frac{1}{n}.
\end{equation}

There is some evidence that the EW method performs similarly, or only slightly worse, compared to alternative weighting methods \citep{Lin2009, Flandoli2011, VanDerFels-Klerx}. However, experts have differing levels of experience or understanding of the literature, or different abilities to quantify their uncertainty. Those that perform better may be assigned a higher weight in the aggregation. This is not necessarily easy though. \cite{Bolger2015} \edit{points out there is still uncertainty involved in the assigning of weights, a problem which the EW method avoids.}





 

 


\subsubsection{The Classical Method}

The Classical Method, or Cooke's Method, (CM) is a mathematical aggregation method that assesses experts based on calibration and informativeness scores \citep{Cooke1988}. These scores are calculated based on a number of seed questions, which are questions the elicitor has answers to, while the experts do not.

The CM calculates weights based on calibration and informativeness. The calibration, $C_i$, takes into account how accurately the expert can specify their uncertainty as probability intervals. If an expert gives a 25\% probability interval, they are well calibrated if the true value falls within the interval 25\% of the time. The informativeness, $I_i$, takes into account how much information the expert has provided. If an expert gives narrow intervals, they are being more informative than if they'd provided wide intervals. 

An additional term, $\alpha_{C_i}$, provides a cutoff for poorly calibrated experts to be excluded from the aggregation. \edit{For experts with calibration $C_i$ under a chosen value, $\alpha_{C_i} = 0$, while $\alpha_{C_i} = 1$ for all other experts. The value of the cutoff is found by choosing the value which maximises the value of the sum of the un-normalised weights, $\sum w_i^*$.} 

The un-normalised weights are

\begin{equation}
 w_i^* = C_i \times I_i \times \alpha_{C_i} 
\end{equation}

which are then normalised by

\begin{equation}
w_i = \frac{ w_i^*}{\sum^n_{i=1}  w_i^*} \text{ .}
\end{equation}

The information score is given by

\begin{equation}
I_i = \sum_{j=1}^r s_j \text{ln}\frac{s_j}{p_j}
\end{equation}

where $s_j$ is the expected proportion of seed questions values that fall within a range, $p_j$ the observed proportion of seed questions values that fall within the range and $r$ the number of ranges elicited for each variable. The calibration score is then given by

\begin{equation}
C_i = P(2qI_i \leq x)
\end{equation}

where $q$ the number of seed questions \citep{Cooke1991}. \edit{The value of $P(2qI_i \leq x)$ is approximated using the $\chi^2$ distribution, with $q-1$ degrees of freedom. The value of $\chi^2_{q-1}(2qI_i)$ is bounded between zero and one.}


The construction of seed questions is an important part of utilizing the CM. \cite{Quigley2018} suggests questions can be either predictions or retrodictions, and either domain or adjacent to the field of expertise. Prediction questions ask the experts about events or quantities that have not yet occurred, and are preferable to retrodiction questions about past events. Domain questions will come from the same field of expertise, and may share the same dimensions as the target questions, while adjacent questions are less related. It is noted that while domain questions are preferable, they may be harder to develop.

Another factor that should be considered is the number of seed questions provided to the experts. A higher number of seed questions allows experts' informativeness and calibration to be better estimated \citep{Eggstaff2013}. It is suggested that there may be an upper bound to the number of seed questions that should be asked, where the addition of extra questions offers no additional improvement to the final distribution. \cite{Clemen2008} suggests the required number of seed questions to ensure the CM performs well is probably greater than 10. \cite{Bolger2015} suggests 10 seed questions isn't enough to determine whether experts are calibrated, and to be able to detect a difference of 0.1 between calibrations would require a sample size of 1546 judgments. A major cost of additional seed questions is the time taken to answer them, which means the trade off between time and estimation accuracy needs to be considered when determining how many seed questions should be provided.

\subsection{Behavioural Aggregation Methods}

Behavioural Aggregation Methods aim to allow a group of experts \secondedit{to} form an aggregated prior through agreement, rather than a mathematical formula. The methods provide a structured communication or discussion framework to assist the experts in forming a consensus. This consensus ensures the resulting distribution has a sensible practical interpretation, as the experts have agreed on it. Mathematical aggregation methods may not offer this without further expert input. A consensus distribution also avoids cases where the experts have major disagreements with the aggregated distributions, as the final distribution would be agreed upon by the group.

Experts may not always be able to form a consensus as a group. In some cases, the elicitation facilitator may be able to assist them in forming a distribution. In other cases, the experts may only be able to aggregate towards a smaller number of distributions than there were experts. For example, if there were two prevalent hypotheses amongst the experts, they may be more comfortable splitting up and creating two separate aggregations. These distributions would then either have to be used separately or aggregated using another method, such as mathematical aggregation. 

The Delphi method was one of the earlier behavioural aggregation methods developed \citep{Dalkey1963}. The method requires some adjustment for prior distribution elicitation, but allows for variations in implementation to best fit the specific analysis to be conducted. \cite{Okoli2004} present a general methodology for implementing the Delphi Method. The Delphi Method is often anonymised between experts, and can be conducted without experts discussing their views in person \citep{EFSA2014}. An alternative which involves a group meeting of all experts is the Sheffield Elicitation Framework (SHELF). 

The Delphi Method and SHELF can not be run simultaneously on the same group of experts, so SHELF was selected as the behavioural aggregation technique used in this study. 

\subsubsection{Sheffield Elicitation Framework}

The Sheffield Elicitation Framework (SHELF) is a method of eliciting probability distributions from a group of experts \citep{Oakley2016}. The SHELF group meeting contains a number of steps, outlined in \cite{Gosling2018}. 

To begin, the experts are trained in the elicitation process. This training includes ensuring they have an understanding of probability and statistics, in which they may have little previous training. It is also recommended that elicitors run the experts through a toy problem, in which they can practise the elicitation and specification of probabilities.

The next stage of the elicitation is for the experts to share information about themselves. The experts are asked about potential vested interests, their expertise, and the sources of evidence they are using to base their judgments on. The information here helps alert the facilitator to potential biases, and strengths and weaknesses of the group, while also reminding the experts of their full range of knowledge. 

The experts are then asked to provide their individual judgments on each of the parameters of interest. It is recommended that the facilitator ask the experts to provide minimum and maximum values, median, and lower and upper quartiles. The order of these values aims to address potential anchoring bias and overconfidence. Other methods, such as the roulette method, can be used instead \citep{Gore1987,Johnson2010}. 

After each expert has specified their individual judgments, distributions are fit using their values. This is usually done using a least squares algorithm, though the exact distribution types to be considered will depend on the parameter of interest.

The individual distributions are then shared with the group, and the experts are encouraged to use them as a basis for creating a group aggregation. The elicitor can use the similarities and differences of the individual distributions to start a conversation, or probe more deeply. Once the experts have formed a consensus, they are then given feedback on the distribution. This can include probabilistic statements from the distribution, or practical interpretations of the parameter. The final aggregated prior is the one resulting from the feedback loop.

It is noted that there may be difficulties in having experts come to an agreement on a single prior distribution given the potential for a large range in different views and experiences. SHELF recommends experts are instead asked to provide a prior for a rational impartial observer (RIO), who has observed their discussion and all of their evidence. This neutral viewpoint aims to help experts to avoid bias, personal investment or interpersonal difficulties. 

\cite{Oakley2016} provides guidance, templates and an R package to assist with the application of SHELF. These documents and templates assist with the management of evidence and definitions for the experts, and the collection of additional information. The R package contains an interactive Shiny application that provides real time feedback of fitted distributions given elicited values. Further applications have been developed for specific studies, for example \cite{Truong2013}.

\section{Comparison of Aggregation Methods}

The aggregation methods were compared using both results from the seed questions and prior distributions of the parameters of interest. Probability intervals were elicited for seed questions in order to calculate weights using the CM. In order to compare aggregation methods, the same questions were asked during a SHELF elicitation to form an additional set of group priors. This allows for CM and SHELF aggregations to be directly compared for the same set of questions, with known answers.

The aggregation methods were applied to the elicitation of priors for use in a sample size calculation for a clinical trial. While many clinical trials use Frequentist methods, Bayesian statistics also offers various approaches \citep{Spiegelhalter1994}.

\subsection{Clinical Trial}

As part of the trial design for a clinical study into a new medical diagnostic test for Motor Neurone Disease (MND), an elicitation was held to develop prior distributions for a number of parameters. The trial aims to determine whether a new experimental diagnostic test (ET), or index test, is able to detect the presence of MND earlier than the current reference test (RT). 


The trial format had previously been designed using frequentist methods, and the purpose of the elicitation was to develop prior distributions for use in determining an appropriate sample size. This paper aims to compare appropriate approaches to eliciting prior distributions \edit{which will be used in sample size calculations and analysis of results}. 

When a patient enters the trial, they will be tested using both the RT and ET. Patients with a negative RT result will be tested using the RT again after six months. Patients who receive a positive RT at the beginning of the trial will leave the trial to receive treatment for the disease.

The patients undergoing these tests are suspected of having the disease. Due to the nature of the disease, it is expected that a patient with MND who receives a negative diagnosis at the start of the trial will deteriorate significantly such that the disease will be detected after a further six months. The aim of the study is to determine whether the ET can detect the presence of MND at the initial time point when the RT would have taken an additional six months.



The results from the trial are to be analysed using McNemar's test \citep{ArmitageP.2008StatisticalResearch}. 



\edit{We consider patients who test positive using the RT at the first time point or after 6 months. We test the null hypothesis that the sensitivities of the RT, and the ET at the first time point, are equal. If s is the number of patients with a positive result from the RT at the first time point and a negative result from the ET, and r is the number of patients that have a negative RT result at the first time point and a positive ET result, then the test statistic is $\frac{(s - r) ^2}{s + r} \sim \chi^2_1$.}

\edit{In order to calculate a sample size using this test, we then need estimates for the value of $r$ and $s$.}

 \edit{In addition to this, a Bayesian analysis of the trial will also be conducted on the recommendation of the trial's grant reviewers. For this, it is planned to calculate the posterior probability of an improvement in diagnosis rate above the MCID, where the MCID is the Minimal Clinically Important Difference, or the improvement the ET would have to offer over the RT in order for clinicians to implement it in their diagnoses.}

\edit{In order to calculate a sample size for the Bayesian analysis, the distribution of the probability a patient with MND receives a positive ET result and negative RT result at the first time point would be required.} The parameters of interest chosen to be elicited were selected to ensure that they could be used in sample size calculations for both a frequentist test and Bayesian analysis.

A model was constructed by breaking the population into subgroups according to patient diagnoses. The terms of the model are summarised in Table~\ref{tab:terms}. For a patient entering the trial, the probability they are positively diagnosed by the RT is denoted by $\eta$. For a patient reaching the second time point of the trial, the probability they are then positively diagnosed by the RT is denoted by $\psi$. The unconditional probability that a patient will receive a positive RT result after six months is then $(1-\eta) \psi$. The probability that a patient does not receive a positive RT diagnosis during the trial will be given by $(1-\eta) (1-\psi)$. 


The patients who received a positive test result from the ET were separated into three groups. The probability a patient diagnosed at the start of the trial using the RT will also receive a positive result from the ET is denoted by $\theta_1$. The probability a patient not diagnosed until after 6 months by the RT receives a positive ET result is denoted by $\theta_2$. The probability a patient who doesn't receive a positive result from the RT at any time point, and receives a positive ET result is denoted by $\theta_3$. 

\edit{The parameters can be combined to calculate the unconditional probabilities for patients. It can be calculated that  the probability of a patient receiving a positive ET and RT at the first time point is $\eta\theta_1$, a positive ET and RT at the second time point is $(1 - \eta)\psi \theta_2$, and a positive ET and negative RT at both time points is $(1-\eta)(1-\psi)\theta_3$. The total probability a patient with receive positive ET is then $\eta\theta_1 + (1 - \eta)\psi \theta_2 + (1-\eta)(1-\psi)\theta_3$.}



\begin{table}
\caption{Model Terms \label{tab:terms}}
\begin{tabular}{c|l}
Parameter & Definition \\ \hline
$\eta$ & \secondedit{P(positive RT at the first time point)} \\
$\psi$ & \secondedit{P(negative RT at the first time point} $\mid$ \secondedit{positive RT at the second time point)} \\
$\theta_1$ & \secondedit{P(positive RT at the first time point} $\mid$ \secondedit{positive ET result)}\\
$\theta_2$ & \secondedit{P(positive RT at the second time point} $\mid$ \secondedit{positive ET result)}\\
$\theta_3$ & \secondedit{P(negative RT for both time points} $\mid$ \secondedit{positive ET result)}
\end{tabular}
\end{table}




\subsection{Elicitation Implementation}
The elicitation for the trial was held on the 21st of March, 2019. \edit{Five parameters were elicited from the experts, $\eta$, $\psi$, $\theta_1$, $\theta_2$, and $\theta_3$. In addition to this, the experts were also asked about ten seed questions. For each parameter and seed question, a median, quartiles, and minimum and maximum were elicited. These questions are provided in the supplementary material.}

Three consultant neurologists and neurophysiologists involved in the development and study of the ET agreed to take part in the elicitation. These experts were selected as they had a strong understanding of the background behind the ET, as well as the current RT methodology. Two of the experts were directly involved in writing the grant application for the trial funding.

The three experts had between eight and twenty-five years of experience of research in MND. Table~\ref{tab:exp} outlines the experts' self assessed knowledge and expertise. All of the experts believed they were at least as knowledgeable about the ET as they were about the RT.

\begin{table}
\caption{Experts' Self Assessed Knowledge and Experience \label{tab:exp}}
\footnotesize
\begin{tabular}{l|ccc}
 & Expert 1 & Expert 2 & Expert 3 \\ \hline
Please rate your knowledge on the disease from\\ \hspace{.5cm} 1 (least) to 5 (most) & 4 & 3.5 & 4.5 \\
Please rate your knowledge on the RT\\\hspace{.5cm} Criteria  from 1 (least) to 5 (most). & 5 & 3.5 & 3 \\
Please rate your knowledge on the ET\\\hspace{.5cm} tests from  1 (least) to 5 (most). & 5 & 4 & 3
\end{tabular}
\end{table}

The experts were also asked to consider their strengths and weakness in terms of MND and providing information during the elicitation. The experts' listed research expertise, background knowledge of the disease and RT, good understanding of electrophysiology, and clinical experience as their strengths. Their self assessed weaknesses included distance from front-line neurology care, a relative lack of on-going lab based research and a shorter amount of practice as a neurologist and clinical neurophysiologist. 

While eliciting information from more experts would have been preferable, the perceived weaknesses of individual experts tended to be covered by the perceived strengths of others. 

\subsubsection{Seed Question Elicitation}

\edit{In order to elicit responses for the seed questions and the parameters as part of the Classical Method, each expert was sent a series of documents. The first document provided information about definitions used throughout the elicitation, and background information about the trial. The second document contained brief questions about the experts' background, and the seed questions. The final document contained instructions for completing the parameter elicitation. These documents are provided in the supplementary materials.}

\edit{Ten seed questions were chosen to use in calculating the weights in the Classical Method. While more seed questions would have provided more information to calculate the weights, it would have taken more time for the experts to complete.}

\edit{The seed questions asked about the rates of MND and the diagnosis of MND under varying circumstances. As the method of diagnosis in the study is novel, there was no data on its diagnosis performance which the experts were not involved in collecting. As such, the questions had to focus on the current methods of diagnosis.}

\edit{The questions were chosen to cover diagnosis over a range of different time periods and locations, and to have the experts consider factors such as the sex of patients and family history. The aim of this was to help remind the experts of their breadth of knowledge to assist in counteracting the availability heuristic, and to have them consider a range of factors which may influence diagnoses.}

\edit{The seed questions were in one of two formats. The first asked the experts for percentages, while the other asked for a number of patients out of 100,000. This was done to assess the experts on a wider range of types of questions, while still allowing them to focus on questions similar to those in the parameter estimation questions.}

\edit{The seed questions were then elicited again using SHELF. Each question was shown to the group, and the anonymised results for the question were provided. The experts then discussed each question, considering the previous responses. The feedback about the seed questions, and aggregated distributions were provided to the experts at a later date.} 

\secondedit{The values elicited from the individual experts and from the SHELF elicitation are provided in the supplementary materials. The supplied values have had the known value subtracted from them, so that the value of zero represents the known answer for all questions.}



\subsubsection{Parameter Elicitation}

Traditionally, the SHELF method conducts individual elicitations immediately before the group elicitation, with the facilitator present to assist the experts. Due to time constraints and expert availability, the SHELF method implemented needed to be modified. Experts completed the individual elicitation stage in their own time, using an online R Shiny application. This allowed the group elicitation session to focus on the discussion stages.

The R Shiny application was developed specifically for the elicitation of parameters for the trial, and was hosted online to allow the experts to complete it in their own time. Experts could input upper and lower limits, quartiles and a median, and the application would fit multiple distributions. The best fitting distribution, as chosen by a least squares algorithm, was presented and the experts could either accept it, or adjust their values or select an alternative distribution. This application is provided in the supplementary materials, and can be accessed here: \url{https://cwilliams.shinyapps.io/shinyelicitation/}.

After this process was completed, the experts were provided with two checks. 

The first check calculated an estimate and probability interval for the total proportion of patients with a \edit{negative RT} at the start of the trial and a positive RT after six months, or $(1 - \eta) \psi$. As this value was not directly elicited, but rather extrapolated from the experts' other distributions, it provided a check to ensure that the elicited values make sense in a new context. If the experts felt this value was incorrect, a number of options were provided as to which elicited values would need to be modified to raise or lower it. 

\begin{figure}
    \centering
\includegraphics[scale=0.5]{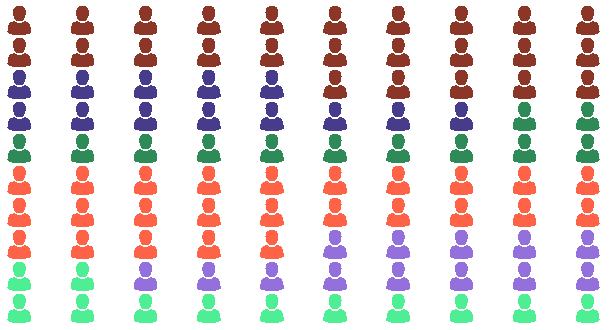}
      \caption{Example elicitation check, providing an example sample of 100 patients using the experts' medians. Positive ETs are coloured in a darker shade, negative ETs are in a lighter shade. Patients with a positive RT in the first round are represented by red. Patients with a positive RT in the second round are represented by purple. Patients with a negative RT from both rounds are represented by green. \edit{The values for each were also provided numerically.}}
    \label{fig:peopleplot}
  \end{figure}

The second check used the median estimates from each parameter to build an example sample of 100 patients. The colour of each patient reflected their test results, and a table summarised further proportions. If this did not represent the experts' beliefs about the results of the trial, then it would identify which parameters needed to be modified further. An example is provided if Figure~\ref{fig:peopleplot}.


\edit{This example figure shows a case where 50 of the patients had a positive ET  result, and are in a darker shade in the top five rows. The remaining 50 patients with a negative ET results are in the lighter shades in the bottom five rows. Of the first 50, 25 had a positive RT test results at the first time point, as shown by their red colouring. The 25 patients in the lighter red shade were those who received a negative ET result, but a positive RT test at the first time point.}


\edit{The patients in Figure~\ref{fig:peopleplot} are determined by the medians of the parameters specified by the experts. Table~\ref{tab:peopleplot} shows how the number of patients in each colour was calculated, for a total sample size of 100. As the medians for each parameter were to two decimal places, the number of patients for each group was also rounded to ensure the total would sum to 100. }

\begin{table}
\caption{Calculations for Figure 1~\ref{fig:peopleplot} \label{tab:peopleplot}}
\begin{tabular}{l|l}
Colour & Number of Patients \\ \hline
Dark Red & $100\eta\theta_1$ \\
Dark Purple & $100\psi(1-\eta)\theta_2$ \\
Dark Green & $100(1-\psi)(1-\eta)\theta_3$ \\
Light Red & $100\eta(1-\theta_1)$ \\
Light Purple & $100\psi(1-\eta)(1-\theta_2)$ \\
Light Green & $100(1-\psi)(1-\eta)(1-\theta_3)$
\end{tabular}
\end{table}

\subsection{Comparison of Aggregation Methods}

The prior distributions elicited for each parameter are challenging to evaluate and compare without results from the trial. Instead, the seed questions have been utilised to compare the methods and evaluate the potential benefit of aggregation. 

A predicted distribution for each seed question was developed using the aggregation methods, and along with the experts' individual distributions, compared to the known answers. The CM distribution was calculated using cross validation, where the nine other seed questions were used to calculate the weights for each of the tenth seed questions. The SHELF and individual expert distributions were provided directly by the experts, and the EW aggregation was calculated from the individual distributions.

\begin{figure}
\centering
\includegraphics[scale=0.45]{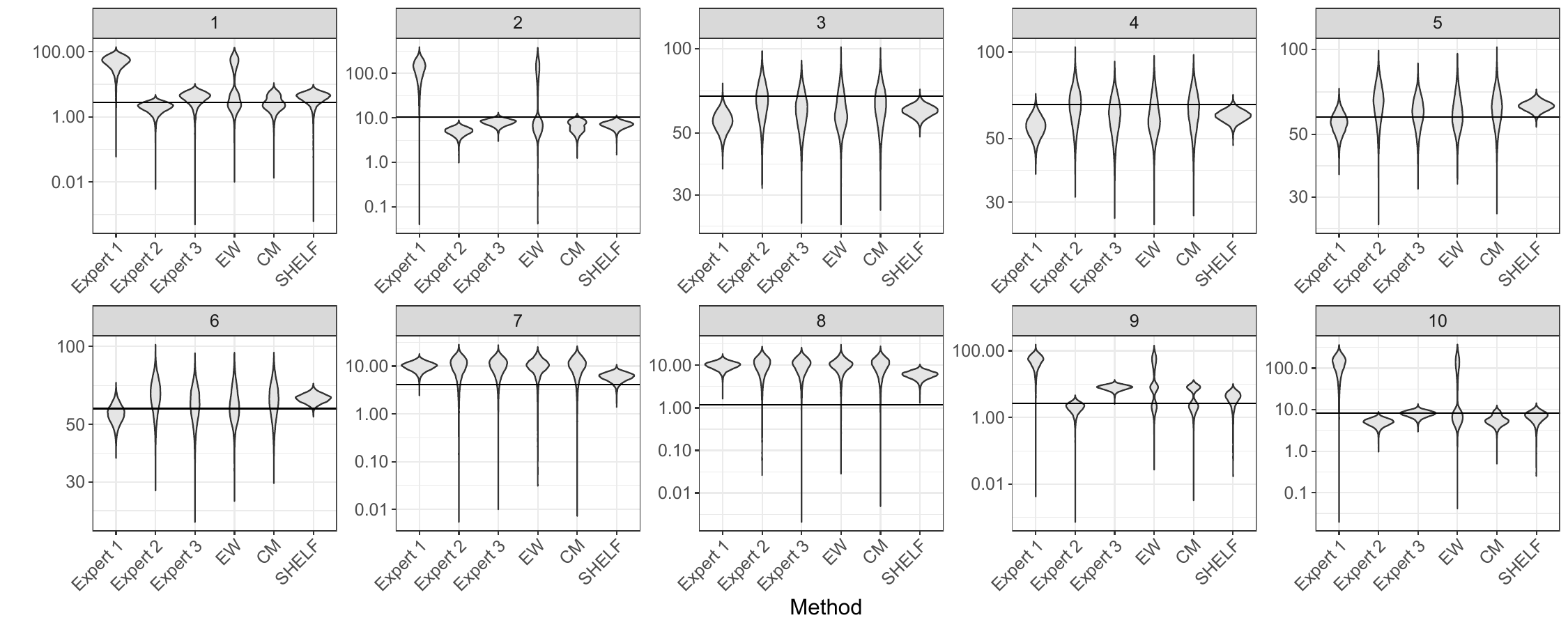}
      \caption{\edit{Violin plots of the individual and aggregated distributions for each seed question. The y axis has been plotted on the log scale for readability, but the quantities were elicited on an untransformed scale.}}
    \label{fig:boxplot}
  \end{figure}

Figure~\ref{fig:boxplot} shows the calculated distribution for each expert and method by seed question. \edit{The violin plots show the density of each aggregation method's distribution over the range of elicited values.}

The three experts provided varying levels of agreement when elicited from individually. In terms of uncertainty, the experts tended to provide distributions with a similar width for each question. For most questions, Experts 2 and 3's distributions tended to be more similar than Expert 1's.

Expert 1 provided notably different responses for Questions 1, 2, 9, and 10, compared to the other experts. This has resulted in EW distributions which are much wider than the other distributions. These wide distributions provide more uncertainty than the other aggregation methods. While this does reflect the initial uncertainty of the experts, as a group a much less uncertain distribution was agreed upon.

\edit{The seed questions 1, 2, 9, and 10 were the four which asked the experts to make judgments of the number of occurrences out of 100,000 people, rather than as a percentage. Expert 1's differences in responses may have been due to the difference in scales, as when they discussed these questions with the other experts in the SHELF elicitation meeting, the views became more aligned.}

The violin plots for the EW distributions cover the entire range of the expert distributions. \edit{However, if there had been any values for which no expert provided density then these values would also be given zero density in the EW distribution}. 

The SHELF distributions tended to be narrower than the other aggregated distributions. This could represent a lower amount of uncertainty for the experts when in a group, or overconfidence. As the SHELF distributions were developed in a group meeting rather than a mathematical aggregation of other elicited distributions, they tend to differ slightly more than EW or CM distributions in comparison to the individual expert distributions.  

The difficulty of the seed questions is also made apparent. Answers for both questions 7 and 8 overestimate the known answer in all distributions. \edit{For these questions, the experts provided a closer midpoint in the SHELF elicitation than individually. They stated that this change away from their original views was due to a further consideration about the time period stated in the question.} While the known answer being in the tail of a distribution is expected occasionally given a well calibrated expert, the cases where all experts provided distributions with the known answer in the very tails or outside the interval suggest the question was difficult. 



Correlations between the errors of the medians for both experts and the group were calculated and have been provided in Table~\ref{tab:seedcor}. A positive correlation suggests that when one expert over or underestimates the true result, the other expert typically does the same. A negative correlation means that when one expert over or underestimates the true result, the other expert typically does the other.

Experts 1 and 2 are negatively correlated and Experts 2 and 3 positively correlated. There does not appear to be any correlation between Experts 1 and 3. \cite{Wilson2017} found that 93\% of the \edit{elicitation} studies tested had correlations between experts of 0.67 or higher, suggesting the correlation found between Expert 2 and 3 is typical. The strong negative correlation between Experts 1 and 2 was found to be less common.

\edit{The differences in the correlations of the errors could have a number of potential causes. Varying experiences, opinions, or ability to specify probabilities could affect the correlations. For example, Experts 1 and 2 reported they were in different career stages, and perceived different levels of expertise, as shown in Table~\ref{tab:exp}. These differences between the two experts could easily lead to them to have conflicting views on their answers. Additionally, the correlation could also be affected by the way the experts provided their probabilities, such as if one consistently tended to overestimate or underestimate compared to the other.}

\begin{table}
\caption{Correlations of Medians' Errors\label{tab:seedcor}}
\begin{tabular}{l|rrrrrr}
         & Expert 1 & Expert 2 & Expert 3 & EW      & CM    & SHELF \\ \hline
Expert 1 & 1        & -0.63    & -0.02    & 0.05    & -0.70 & -0.14 \\
Expert 2 & -0.63    & 1        & 0.68     & 0.62    & 0.25  & 0.80 \\
Expert 3 & -0.02    & 0.68     & 1        & 0.99    & -0.50 & 0.86 \\
EW       & 0.05     & 0.62     & 0.99     & 1       & -0.57 & 0.83 \\
CM       &-0.70     & 0.25     & -0.50    & -0.57   & 1     & -0.14 \\
SHELF    & -0.14    & 0.80     & 0.86     & 0.83    & -0.14 & 1 
\end{tabular}
\end{table}

Three proper scoring rules were selected to evaluate the individual responses and aggregated distributions. The Brier, Logarithmic and Quadratic scores each reflect the informativeness and calibration of the experts \citep{Winkler1996,JamesE.MathesonandRobertL.Winkler1976}. For the scoring rules that require a density function, a normal or log normal distribution was fit to each expert's quartiles. This appeared reasonable given the mostly symmetric violin plots of the provided values. The resulting scores have been provided in Table~\ref{tab:scores}. 

The logarithmic scoring rule calculates the negative of the log value of the density at the location of the known value. At the value $r$,

\begin{equation}
    L(r) = - \ln f(r)
\end{equation}

As only quartile boundaries were elicited and calculated, in order to determine the value of the density at a given point a distribution was fit to each question. Using least squares, the normal or log-normal distribution that best fit the provided quartiles was selected.

The second method used was a Brier Score. This calculates the squared error between the known value, and the distribution's central estimate, in this case the median. For the known value $r$, and an expert's median $m$,

\begin{equation}
    B(r) = (m - r)^2
\end{equation}


The Quadratic Score calculates the integral of the distribution squared, and subtracts it from the value of the density at the location of the known value multiplied by two.

\begin{equation}
    Q(r) = 2f(r) - \int^{\infty}_{-\infty}(f(\theta))^2 d\theta
\end{equation}

An expert who is well calibrated and informative will receive a low value for their Logarithmic and Brier Scores, and a high value for the Quadratic Score. While both calibration and informativeness are reflected in each score, it is expected the Brier score will more strongly reflect the informativeness of the experts, the Logarithmic score will more strongly reflect the calibration of the experts, and the Quadratic score will provide a more balanced position between them.

\begin{table}
\caption{Cross validation prediction scores \label{tab:scores}}
\centering
\begin{tabular}{l|llll}
& Brier & Logarithmic &  Quadratic\\ \hline
Expert 1 & 2939.8 & 48.5 & 0.18 \\
Expert 2 & 26.9 & 45.9 & 0.19 \\
Expert 3 & 24.7 &  40.2 & 0.29 \\
EW & 26.6 & 37.3 & 0.51 \\
CM & 14.7 & 36.3 & 0.50 \\
SHELF & 18.8 & 34.9 & 0.64
\end{tabular}
\end{table}

The Brier Score has been used to measure the accuracy of the \edit{medians of the aggregated priors}. The median is often used as a best single point estimate of an expert, and so a lower Brier score represents a more accurate expert. The CM aggregation resulted in the lowest Brier score. Both the CM and SHELF offered an improvement over any single expert, demonstrating an increase in accuracy due to aggregation. While the EW Brier score was higher than the other aggregation methods', it performed much better than Expert 1 and similarly to Expert 2. 

The Logarithmic Score suggests that all three aggregation method performed better than the individual experts. Both the CM and EW performed similarly to each other, though both were outperformed by SHELF. All three scores were quite close in comparison to the individual experts, and so there appears to be an improvement provided by aggregating, regardless of method.

The Quadratic Score showed the largest difference between aggregated distributions and the experts. While EW and CM performed essentially equally well, the SHELF aggregation performed better than both. Expert 3 outperformed the other two experts, who received very similar scores. 

Expert 3 outperformed the other two experts on all scores, while Expert 1 performed the least well. This is not to suggest that Expert 1's views were unnecessary or irrelevant, just that there is a discrepancy between their best estimates, their ability to quantify uncertainty and the true answer. It could be their distributions accurately represent their beliefs, which do not correspond with the true answer. Alternatively, they may be overestimating their certainty, or have misinterpreted the questions.

\subsection{Elicitation Results}

\begin{figure}
    \centering
\begin{minipage}[t]{0.45\textwidth}
\vspace{0pt}
\includegraphics[scale=0.35]{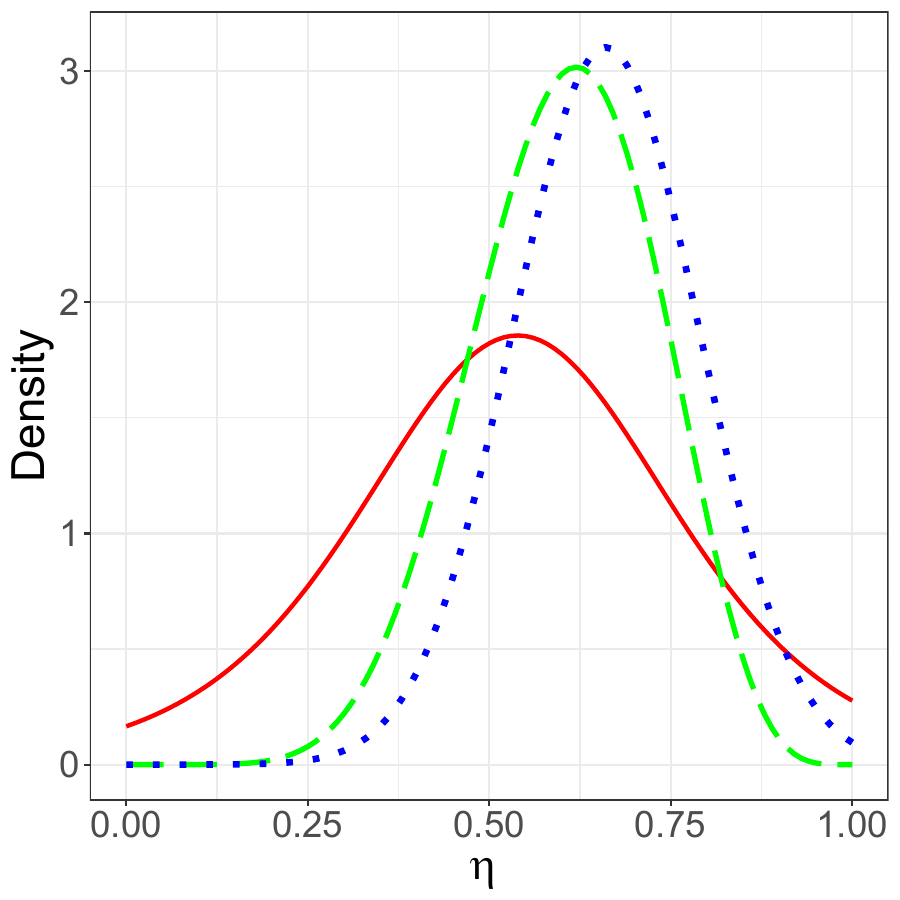} \includegraphics[scale=0.35]{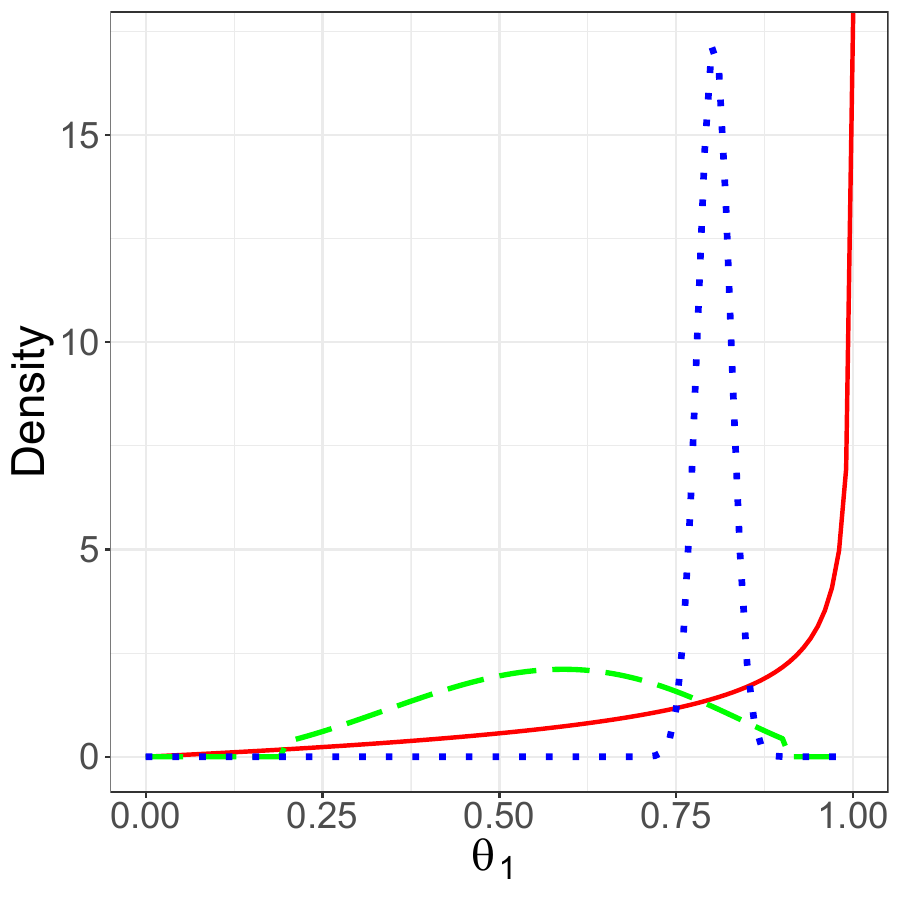} \includegraphics[scale=0.35]{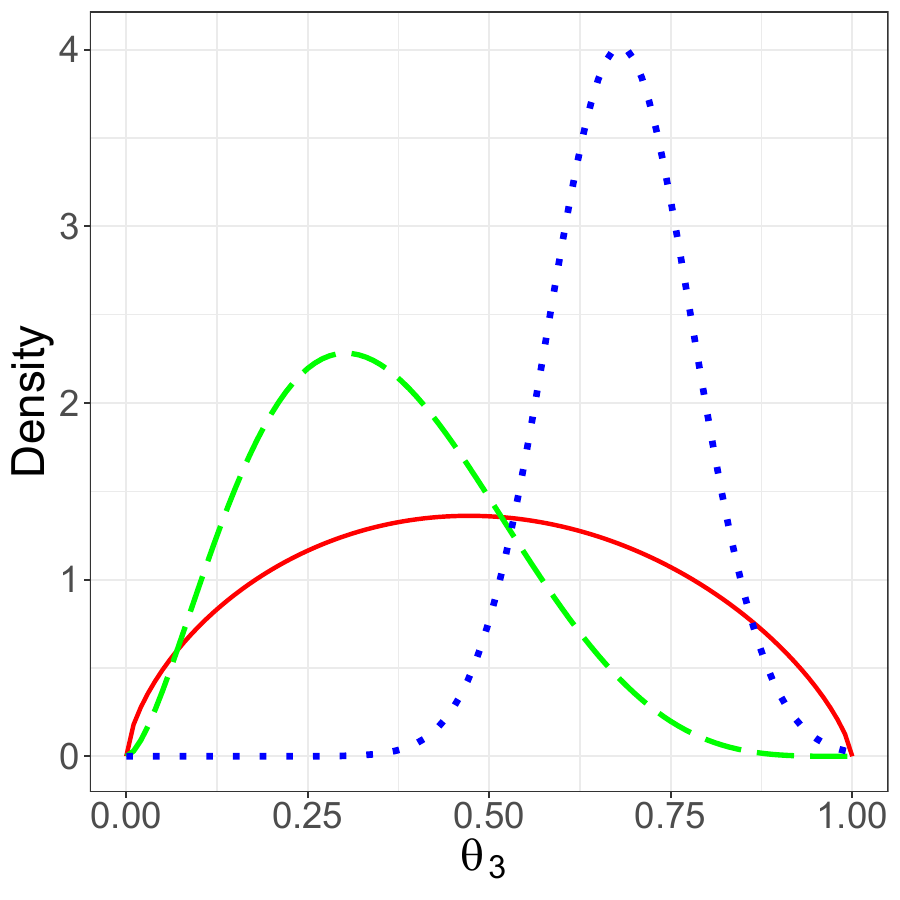}      
\end{minipage}
\begin{minipage}[t]{0.45\textwidth}
\vspace{0pt}
\includegraphics[scale=0.35]{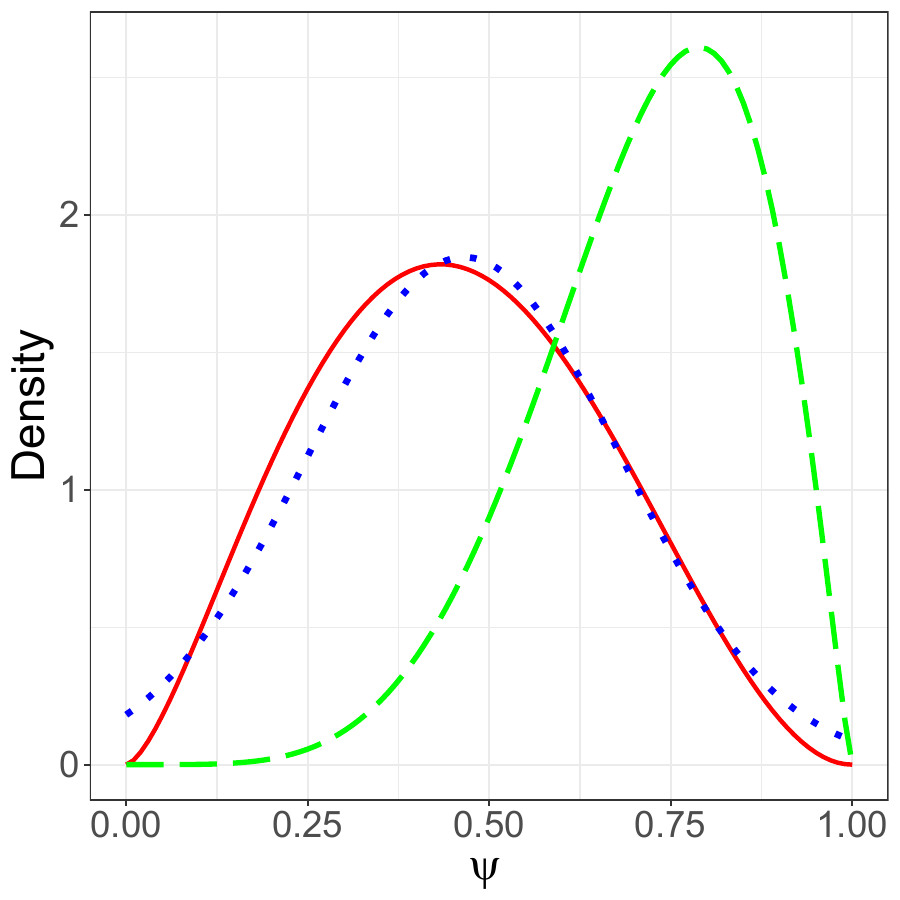}   \includegraphics[scale=0.35]{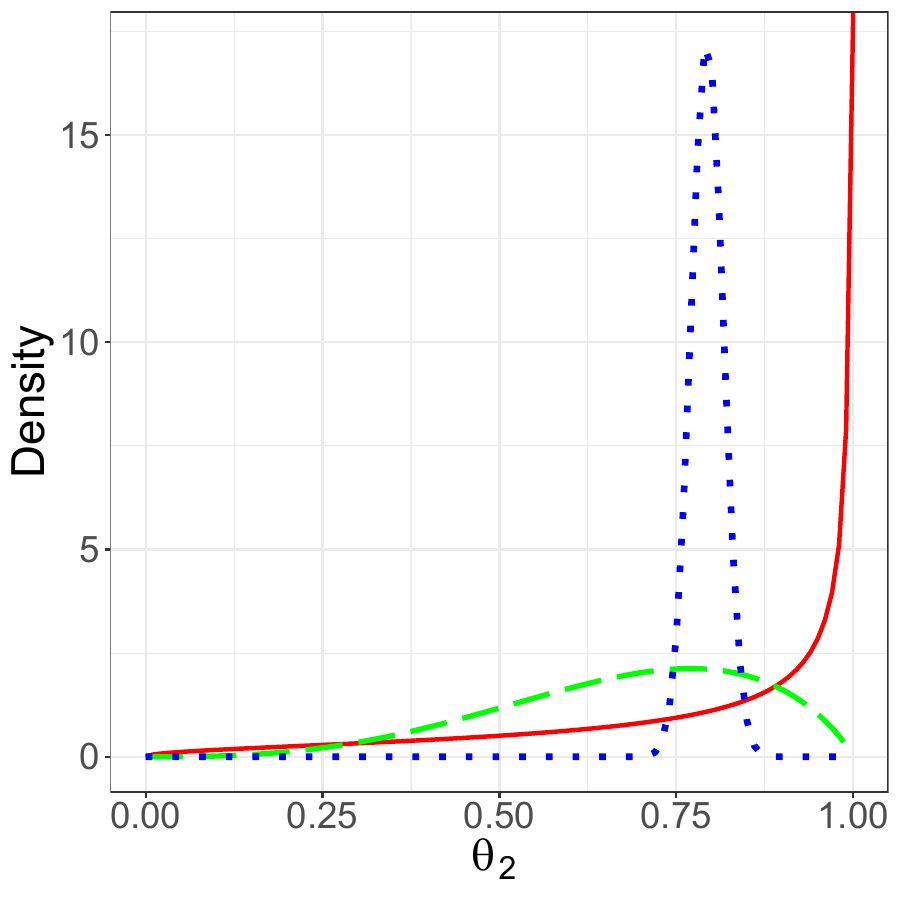}
\end{minipage}
      \caption{Expert prior distributions from Expert 1 (red, solid line), 2 (green, dashed line), and 3 (blue, dotted line).}
    \label{fig:1}
  \end{figure}

The initial individual prior distributions for each parameter are given in Figure~\ref{fig:1}. 

The agreement between experts varied by parameter. Some, such as $\eta$ and $\psi$, have a pair of experts providing very similar distributions, while others such as $\theta_3$ resulted in three unique distributions. The experts in agreement changed between parameters, with Experts 2 and 3 providing similar prior distributions for $\eta$ and Experts 1 and 3 for $\psi$. This suggests the similarities aren't due to a single pair of experts having a similar background or knowledge base, but rather an agreement of viewpoints.


While the distributions varied between individual experts, each provided a similar distribution for $\theta_1$ and $\theta_2$. This implies the probability a patient would receive a positive ET result would be the same unconditional on the results from the RT. The interpretation here may be that the experts view the two tests to be measuring different biological factors which are independent of each other. The viewpoint that both distributions would be the same, regardless of their view of the actual distribution, was an important piece of information which demonstrated an agreement by the experts.  

The $\theta_3$ parameter appeared to have the least agreement between the experts. Expert 1 gave a vague, centred prior, Expert 2 a prior with more weight towards values less than 0.5, and Expert 3 a prior with more weight towards values larger than 0.5. This lack of consensus suggested this parameter was harder to form an opinion on. Given the parameter represented a positive ET and negative RTs at both time points, it made sense that it would cover many possibilities. Within $\theta_3$, experts needed to consider the false positive rate of the ET, the cases where the RT was unable to detect a form of MND which the ET could, cases where the RT may have provided a false negative, and other scenarios.  

There were also further patterns with the experts. Expert 2 tended to be more conservative with their estimates of the ET, and more confident in the RT's effectiveness. Expert 1 appeared to be more conservative about the current RT's effectiveness, and more confident about the ET's ability to detect MND. Expert 1 clearly believed the ET would perform much better than the RT in comparison to Expert 2.

Expert 3 often provided narrower distributions, suggesting less uncertainty in their beliefs or an overconfidence in their estimates. While distributions with less uncertainty are more informative priors, this is only a benefit if the expert is well calibrated and the distribution is still accurately specifying this uncertainty.











\begin{figure}
    \centering
\begin{minipage}[t]{0.45\textwidth}
\vspace{0pt}
\includegraphics[scale=0.35]{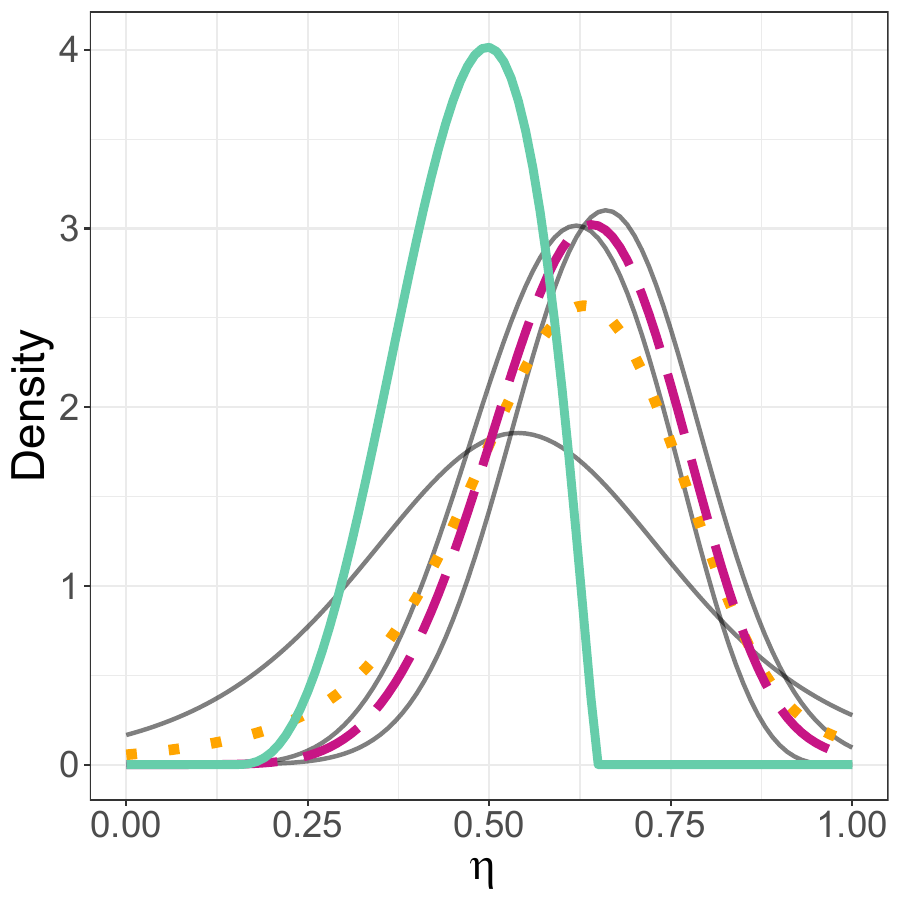} \includegraphics[scale=0.35]{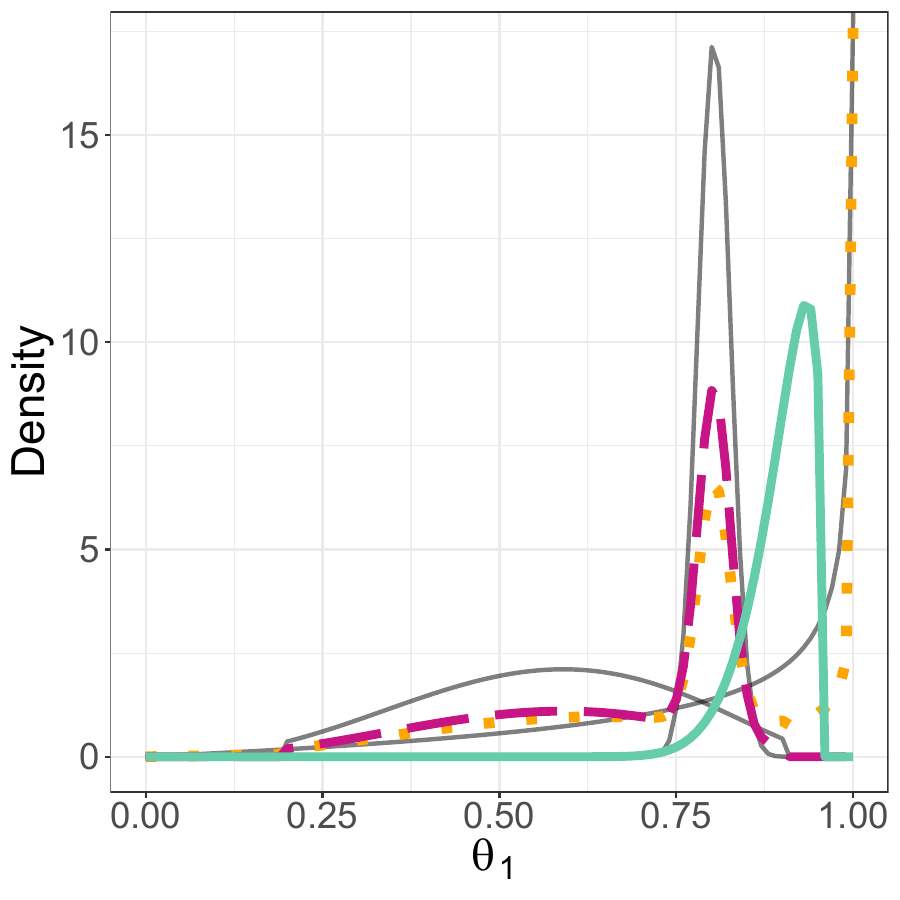} \includegraphics[scale=0.35]{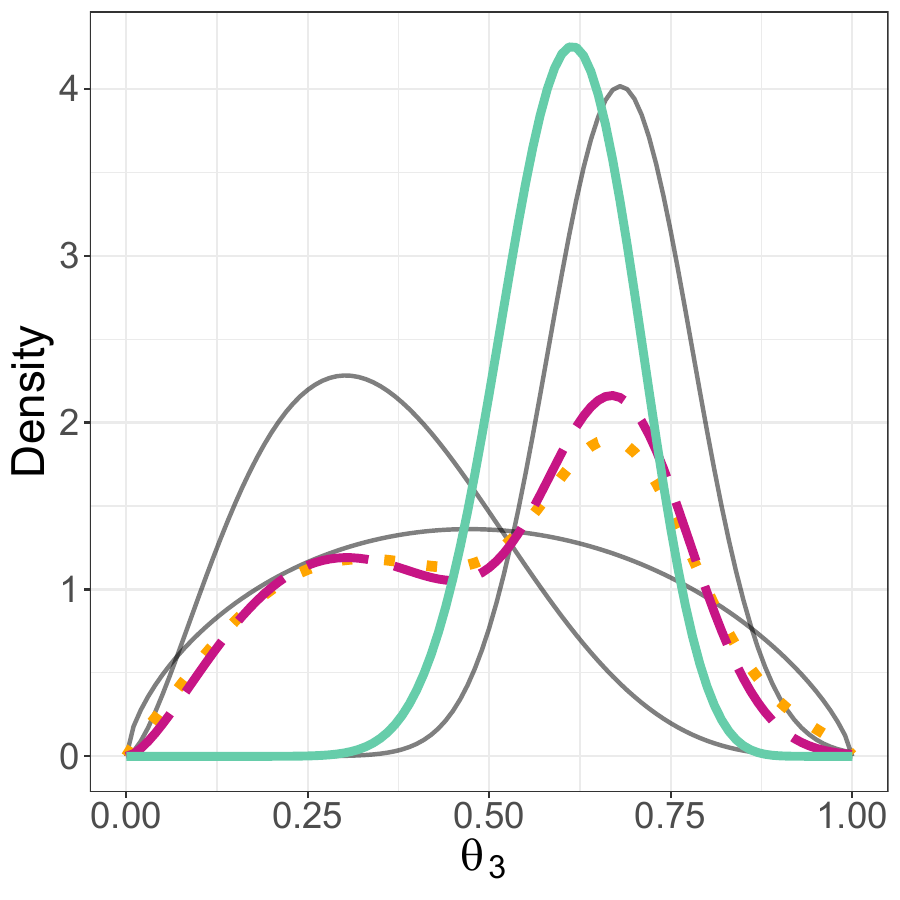}      
\end{minipage}
\begin{minipage}[t]{0.45\textwidth}
\vspace{0pt}
\includegraphics[scale=0.35]{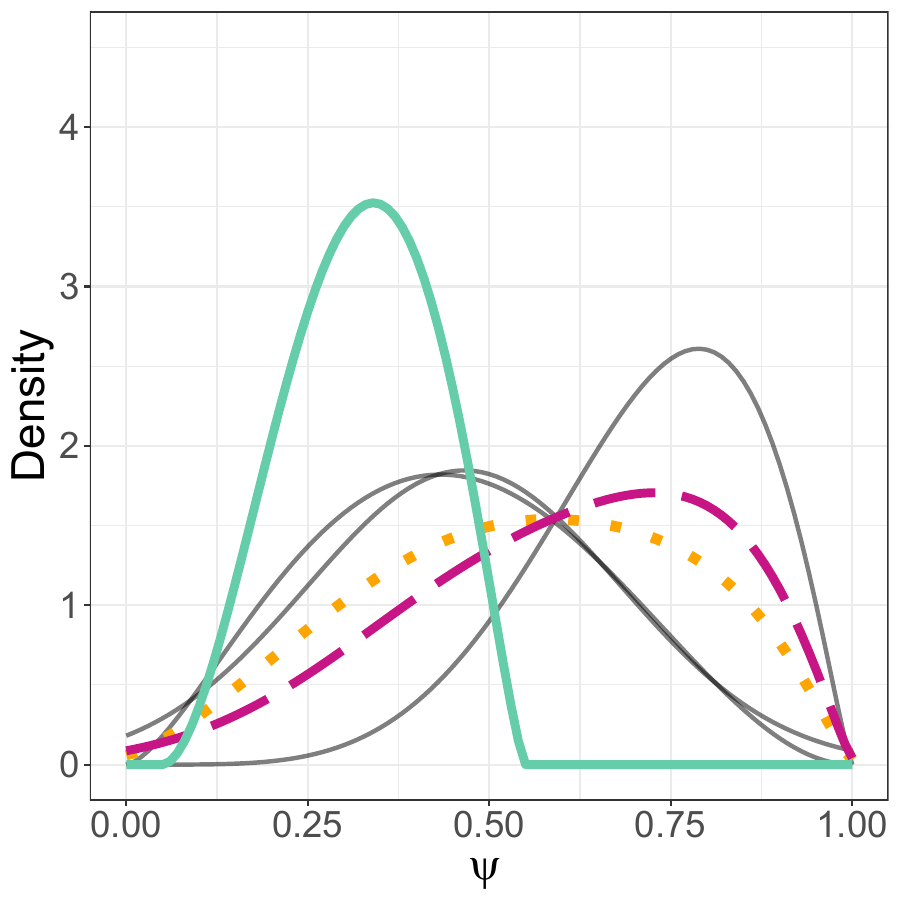}   \includegraphics[scale=0.35]{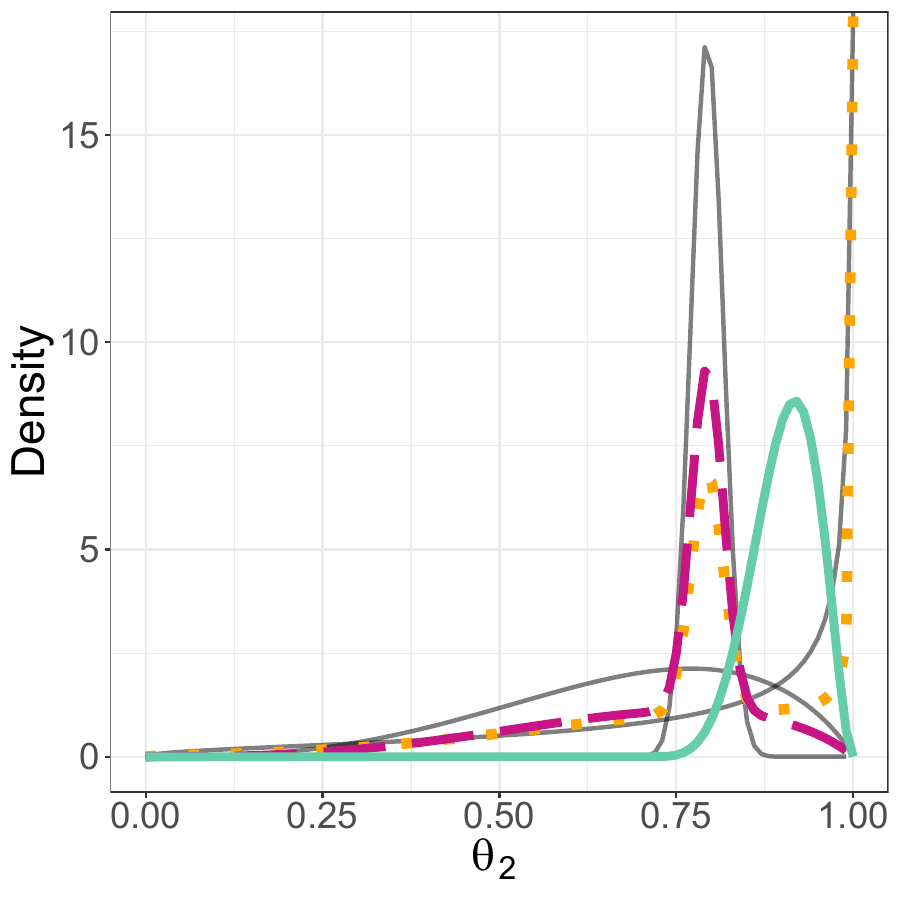}
\end{minipage}
      \caption{Aggregated prior distributions from EW (orange, dotted line), CM (red, dashed line), and SHELF (green, solid line). \secondedit{Prior distributions from individual experts} are also provided (black, thin line).}
    \label{fig:2}
  \end{figure}

Aggregated prior distributions were calculated for each parameter, and are provided in Figure~\ref{fig:2}. The coloured curves represent the aggregated priors, while the black curves represent the individual experts. 

As mathematical aggregation methods, both the EW and CM distributions are linear combinations of the three expert distributions. These two distributions tend to more closely match the original expert distributions than the SHELF distribution. In most cases, the SHELF distributions did not closely match any of the individual experts' priors. While in some cases, for $\theta_1$ and $\theta_2$, SHELF appeared to form a consensus between all experts' priors, for $\eta$ and $\psi$ the experts appeared to come to a slightly different conclusion as a group. For $\theta_3$, the SHELF prior was quite similar in shape and location to one of the expert's priors.

The CM only gave weight to two experts, which is especially noticeable in the upper tails of $\theta_1$ and $\theta_2$. Expert 1's seed question results suggested a low level of calibration, and so they weren't included in the final CM distributions. The CM has assigned weights of 0.52 and 0.48 to Experts 2 and 3 respectively, while the EW method assigned each expert a weight of $\frac{1}{3}$. Despite this difference in weights, the prior distributions created by these aggregation methods were all relatively similar. 

\edit{It should be noted that should the weightings have excluded seed questions 1, 2, 9, and 10 for which Expert 1's responses were quite different to the other experts, the weightings of the three experts would have been 0.45, 0.26, and 0.29 respectively. This set of four seed questions were the ones on a scale of 100,000 people, while the other six questions asked about percentages.}

The SHELF distributions tended to have lower variances than the mathematical aggregation methods. This was also the case in the seed questions, suggesting that the experts behaved similarly in both sets of questions. 

EW lead to bimodal distributions for each of the $\theta$ parameters. \edit{In terms of a single expert's views, it seems implausible that the parameter distributions would be bimodal.} For example, values of around 0.8 and 1 for $\theta_2$ have been assigned a high probability, while values around 0.9 have quite low probabilities. Conceptually, there does not seem to be a plausible reason why the test would not likely detect the disease with a probability of 0.9 if it would likely detect with a probability of 0.8 or 1. \edit{From a decision maker's perspective, this could be used as a way to account for experts who do not agree on a single distribution. However, when the experts can come to an agreement using another method, a unimodal distribution allows for a more practically justifiable distribution.}

The prior distributions for $\theta_3$ demonstrate one advantage of the SHELF method. As there was little agreement between experts, the mathematical aggregation distributions remain vague. It is also questionable as to whether a bimodal distribution is appropriate for this parameter. By allowing the experts to discuss the problem as a group, the SHELF method has resulted in a uni-modal, narrower distribution.

As previously noted, the experts gave very similar distributions for both $\theta_1$ and $\theta_2$. This pattern is again present in the aggregated distributions, with each method providing very similar distributions for both parameters. If an aggregation method had provided different priors for the two parameters, it may have suggested the method was not giving a result consistent with the experts' knowledge.

\section{Discussion}


The benefit of using an aggregation method is demonstrated in Table~\ref{tab:scores}. Should a single expert be elicited from alone, the suitability of their prior distributions relies on both their knowledge of the field and their ability to specify probabilities. An expert who does not perform well in the elicitation may result in distributions that do not reflect the final outcome, yet that same expert can be included within a group which performs better than any individual expert. 

Furthermore, during a single elicitation with a single expert it is hard to judge their ability, making it difficult to determine whether a well calibrated expert has been selected or not. Should a single expert be elicited from, the decision maker would have no knowledge of whether the resulting priors are accurate. While a score could be calculated for an individual expert, as was done in Table~\ref{tab:scores}, the resulting score is only contextualized by the scores of the other experts.

Discussions involving a range of people are likely to reduce the uncertainty surrounding their individual abilities to specify probabilities, and knowledge of the domain of interest. Ideally the range of experts would also account for implicit biases, such as those relating to personal interests in the outcome of a trial, but these cannot always be avoided. The group setting used in SHELF facilitates this type of discussion, though does have some potential downsides. As seen in the seed question example, SHELF resulted in narrower distributions than the other aggregation methods. While this could simply reflect a decrease in uncertainty, if could also represent an overconfidence in the accuracy of their views. 

\edit{For the parameters $\theta_1$, $\theta_2$ and $\theta_3$, the SHELF distributions lie between the three experts' distributions. The experts with lower estimates for $\theta_1$ and $\theta_2$ agreed to increase the estimate after the other expert explained their reasoning, specifically around the variety of different types of MND.}

\edit{The parameters $\eta$ and $\psi$ both had SHELF distributions with more density lower than the individual experts. During their discussions, the experts felt that they needed to be stricter with the upper limit of their distribution after considering published data on the RT. While these distributions have narrower ranges, they still have modes in regions where the experts placed high density.}

\edit{It is recommended that a different form of judgement is used in the group elicitation compared to the individual elicitation stage of SHELF. This was not done for this elicitation to save time in re-training the experts, however it was ensured that multiple forms of judgements were used in the seed and trial elicitation questions.}

The CM and SHELF are both widely used aggregation techniques, and the comparison of seed questions shows they both perform well. Both methods consistently outperformed any individual expert, suggesting the use of either would result in a more informative and better calibrated prior distribution. EW aggregation offers a quicker and simpler method of aggregation, though there is less of an improvement over individual experts. Still, without seed questions with which to judge experts to select the ones that performed best, using an EW aggregated prior would perform better than randomly selecting an expert.

Previous studies comparing mathematical aggregations and individual experts tended to find similar results as found here. \cite{Cooke2008} found the CM outperformed EW and the best expert. \cite{Ganguly2014} showed that over a group of 48 datasets, EW's error was 3\% higher than the CM, while the majority of the time the CM was better calibrated than the \edit{EW}.  \cite{Flandoli2011} suggested that any of the aggregation methods would perform better than selecting a single expert at random, and that none of the aggregation methods tested ever performed worse than the single best expert. Likewise, \cite{Lin2009} finds aggregation performs better than a single expert, and that the CM and EW perform similarly. 
\edit{However, as \cite{Hammitt2013} found the best expert performed better than either aggregation method it appears that whether an expert can outperform the aggregation depends on the specific elicitation results.} In this case, the CM still outperformed EW. While these comparisons looked at various aggregation methods, the comparison of SHELF to the CM has not been previously analysed. 

The CM performed better than SHELF in the Brier score, suggesting a more informative prior, while SHELF received better results under the Logarithmic and Quadratic scores, suggesting a better calibrated prior. In terms of scores, the performance of both the methods appears strong. The benefits of one over the other may therefore lie more in the specifics of their applications.

When eliciting from a group of experts using the CM, each expert can complete the elicitation in their own time. For a SHELF elicitation however, each expert is required to attend the group elicitation meeting. As with this case study, and many other elicitations, there can be difficulty in organising a group elicitation. Experts are usually involved in many projects, and finding a suitable time in which there are multiple hours of availability for all experts, and the elicitation coordinators, can be difficult. 

The resulting aggregated distribution from the SHELF meeting does have the advantage of being directly specified by the experts. This suggests that there will be a practical interpretation of the distribution which the experts have agreed upon. A CM aggregation does not necessarily have this benefit as, being a mathematical aggregation, it results in a weighted average of the individual experts. This can lead to multi-modal distributions, which may not be appropriate for a given parameter. 


Table~\ref{tab:seedcor} shows correlations between the errors of the median values of the seed questions of all experts and aggregation methods. EW and CM had a negative correlation, while EW and SHELF had a strong positive correlation. As the CM only selected two experts to include, the negative correlation between EW and CM may likely be due to the effect of the additional expert's distribution. The correlation between EW and SHELF suggest the experts had chosen SHELF medians that were close to the EW medians. 

Expert 3 was very highly correlated with the EW method. It appears this was the case as Expert 3's median was always between the other two, ensuring that when the three expert distributions were summed that the aggregated median fell closest to Expert 3's. 

SHELF was highly correlated with both Expert 2 and 3, suggesting the experts as a group agreed on a viewpoint more similar to those experts. It also suggests that no one expert dominated the discussion, and pulled the aggregated distribution towards solely their own.

While mathematical aggregation methods can lead to strangely shaped mixture distributions, there are benefits to their use. By specifying an aggregation rule before the elicitation, it removes potential inter-group biases that may occur in a group elicitation, or perceptions of bias due to the aggregation method chosen. The three experts reported significantly different distributions for the $\theta_3$ parameter, yet as a group came to a compromise distribution between the three. Had one of the experts not agreed with the rest of the group, or had been overly forceful with their views, it could have been the case where no consensus, or a very skewed consensus, was formed.  

While it appeared Expert 1 was more optimistic than Expert 2 about the effectiveness of the ET, it isn't necessarily the case that one is correct and the other isn't. They could both be presenting different views representative of the field by taking different experiences or knowledge into account. These different slices of the current state of knowledge would both be valid representations of different perspectives, and so the aggregation of them would better represent a more overarching view. 

Alternatively, one could be overly optimistic, or the other overly pessimistic, or both. With the aim of providing an neutral prior, aggregating over the experts would help reduce the effects of any bias. 

Experts who have been directly involved in the development of a new medical test would likely hold positive beliefs about the effectiveness of their test in order to put it forward to trial. This potential conflict of interest could lead to an overly optimistic prior, or at least the perception of one. While this could be avoided by eliciting from experts not directly involved in the trial, for novel methodologies the trial designers may be the only ones with the relevant knowledge.    

It was also noted that some differences between the individual elicitations and group elicitations may have been due to the presence of the facilitators. The experts completed the individual elicitation alone, and while they could contact the facilitators if they needed any clarification, they did mention that the further explanation available in the group meeting was useful. This suggests that while the modifications to SHELF to allow the first half to be conducted remotely were useful for time management, they may have produced less precise estimates than they otherwise would have. The final elicitation for the group prior did have a facilitator present to clarify any issues the experts had, and so it is expected the effect on the group prior should be minimal. This may have assisted the SHELF method to perform better than other aggregations, though the effect is hard to measure.

While only three experts were available for the elicitation, it is anticipated a further elicitation will investigate the views of experts not directly involved in the trial. This will allow for comparisons between different groups of experts with different levels of interest in the trial, and for further comparisons between aggregation methods.

\section{Conclusion}

There appears to be strong benefits to aggregating the judgements of multiple experts, rather than using just a single expert to provide information for the construction of prior distributions. Under three proper scoring rules, the CM and SHELF outperformed all individual experts. While the EW method was outperformed by a single expert, in a practical setting the elicitors would not know which expert would outperform the aggregation.

Based on the results from the seed questions, there was some evidence that the SHELF outperformed both an EW aggregation and the CM. It should be noted, however, that SHELF requires the greatest work for the experts, as they are required to provide their priors twice; once individually and once as a group. There can also be difficulties with scheduling an appropriately long elicitation meeting. On the other hand, SHELF offers additional benefits over purely mathematical aggregation methods by bringing all the experts together, and allowing them a structure to discuss and construct a prior distribution. Where possible, we suggest this trade off is worth the additional time requirements for a potentially more accurate prior, and additional qualitative information from the SHELF meeting.

\section{Acknowledgments}
The authors would like to thank Dr Mark Baker, Dr Tim Williams and Dr Stephan Jaiser for their assistance in the elicitation.

\newpage
\renewcommand\refname{Bibliography}

\bibliographystyle{plainnat} 
\bibliography{main} 

\end{document}